\documentclass[aps,prl,amssymb,twocolumn,superscriptaddress,showpacs]{revtex4}
\pdfoutput=1
\usepackage{graphicx}
\usepackage{dcolumn}
\usepackage{bm}

\newcommand{\Tr}{{\rm Tr}}
\newcommand{\beq}{\begin{equation}}
\newcommand{\eeq}{\end{equation}}
\newcommand{\avg}[1]{\left< #1 \right>}
\newcommand{\bra}[1]{\big<#1|}
\newcommand{\ket}[1]{|#1\big>}
\newcommand{\barr}{\begin{eqnarray}}
\newcommand{\earr}{\end{eqnarray}}
\newcommand{\Ord}[1]{{\cal O}\left( #1\right)}

\def\cD{{\cal D}}
\def\cZ{{\cal Z}}
\def\pvint{-\!\!\!\!\!\!\int}

\begin{document}

\author{P. Facchi}
\affiliation{Dipartimento di Matematica, Universit\`a di Bari,
        I-70125  Bari, Italy}
\affiliation{INFN, Sezione di Bari, I-70126 Bari, Italy}
\author{U. Marzolino}
\affiliation{Dipartimento di Fisica, Universit\`{a} di Roma ``La Sapienza", Piazzale Aldo Moro 2, 00185 Roma, Italy}
\author{G. Parisi}
\affiliation{Dipartimento di Fisica, Universit\`{a} di Roma ``La Sapienza", Piazzale Aldo Moro 2, 00185 Roma, Italy}
\affiliation{Centre for Statistical Mechanics and Complexity (SMC), CNR-INFM, 00185 Roma,
Italy\\
INFN, Sezione di Roma,  00185 Roma,
Italy}
\author{S. Pascazio} \affiliation{Dipartimento di Fisica,
Universit\`a di Bari,
        I-70126  Bari, Italy}
\affiliation{INFN, Sezione di Bari, I-70126 Bari, Italy}
\author{A. Scardicchio} \affiliation{Princeton Center for Theoretical Physics\\ and Physics Department,
Princeton University, Princeton, 08542 NJ, USA}
\affiliation{MECENAS, Universit\`a Federico II di Napoli, Via
Mezzocannone 8, I-80134 Napoli, Italy}

\title{Phase transitions of bipartite entanglement}

\date{\today}

\begin{abstract}
We study a random matrix model for the statistical properties of the
purity of a bipartite quantum system at a finite (fictitious)
temperature. This enables us to write the generating function for
the cumulants, for both balanced and unbalanced bipartitions. It
also unveils an unexpected feature of the system, namely the
existence of two phase transitions, characterized by different
spectra of the density matrices. One of the critical phases is
described by the statistical mechanics of random surfaces, the other is
a second-order phase transition.
\end{abstract}

\pacs{03.67.Mn, 03.65.Ud, 68.35.Rh}

\maketitle

The bipartite entanglement of small quantum systems (such as a pair
of qubits) can be given a quantitative characterization in terms of
several physically equivalent measures, such as entropy and
concurrence \cite{woot}. The problem becomes more complicated for
larger systems and/or higher dimensional qudits \cite{multipart}.
The interest of characterizing entanglement for these systems is
twofold: on one hand, it has fascinating links with complexity
\cite{parisi} and a related definition of multipartite entanglement
\cite{FFPP}; on the other hand, it has applications in quantum
information and related fields of investigation \cite{nielsen}.

In this Letter we intend to characterize the statistics of the
entanglement of a large quantum system. We shall tackle this problem by
studying a random matrix model that describes the statistical
properties of the purity of a bipartite quantum system. In the
context of quantum information this model was introduced in
\cite{aaa,page} in order to describe the statistics of the eigenvalues of
the reduced density matrix of a subsystem and extract the first
moments of some quantities of interest, like the entanglement
entropy or the purity. We will obtain the exact generating function
of the purity in the limit of large space dimension (large $N$ in
the matrix model) and will connect the entropy with the volume of
the manifolds with constant purity (iso-purity manifolds). We will
also show that the matrix model undergoes two phase transitions, one
at a negative and one at a positive (fictitious) temperature. The
phase transition at negative temperature will be paralleled to
another one, that is well known in the study of random matrix models and conformal field theory literature \cite{Di Francesco:1993nw}. We notice that techniques
related to those presented in this Letter have been recently
employed \cite{Majumdar07} to analyze the statistics of the lowest eigenvalue
of the reduced density matrix.

Consider a bipartite system in the Hilbert space
$\mathcal{H}=\mathcal{H}_A\otimes\mathcal{H}_B$, with
$\dim\mathcal{H}_A=N \leq \dim\mathcal{H}_B=M$. Assume that the
system is in a pure state $\ket{\psi}\in\mathcal{H}$. The reduced
density matrix of subsystem $A$ reads
\begin{equation}
\rho_A=\Tr_B\ket{\psi}\bra{\psi}
\end{equation}
and is a hermitian, positive, unit-trace $N\times N$ matrix. Its
purity
\begin{equation}
\pi_{AB}=\Tr_A\rho_A^2 \in [1/N,1]
\end{equation}
is a good measure of the entanglement between the two subsystems:
its minimum is attained when all the eigenvalues are $=1/N$
(completely mixed state, maximal entanglement between the two
bipartitions), while its maximum detects a factorized state (no
entanglement).
We consider a typical
pure state $\ket{\psi}$ \cite{aaa,page}, sampled according to the
unique, unitarily invariant Haar measure. The significance of this measure
can be understood in the following way: consider a state vector
$\ket{\psi_0}$ and let consider a unitary transformation $\ket{\psi}=U
\ket{\psi_0}$. In the least set of assumptions on $U$, the measure can be chosen randomly in a unique way. The final state
$\ket{\psi}$ will hence be distributed according to the Haar measure mentioned above
(indipendently of $\ket{\psi_0}$). Notice the analogy with the
maximum entropy argument in classical statistical mechanics. By
tracing over subsystem $B$, this measure translates into the measure
over the space of Hermitian, positive matrices of unit trace
\cite{aaa,page}
\begin{eqnarray}
d\mu(\rho_A) &=& \cD\rho_A (\det \rho_A)^{M-N} \delta(1-\Tr \rho_A), \nonumber \\
&=& d^N\lambda \prod_{i<j}(\lambda_i-\lambda_j)^2
 \prod_\ell \lambda_\ell^{\mu N} \delta (1-\sum_k \lambda_k ),
\label{eq:measure}
\end{eqnarray}
where $\lambda_k$ are the positive eigenvalues of $\rho_A$ (Schmidt coefficients), we dropped the volume of the $SU(N)$ group  (which is
irrelevant for our purposes) and $\mu N\equiv M-N$ is the difference between the dimensions of the Hilbert
spaces $\mathcal{H}_A$ and $\mathcal{H}_B$.

We will consider the statistical properties of the rescaled quantity
\begin{equation}
\label{RA}
R=R_{AB}=N^3 \pi_{AB}.
\end{equation}
The moments of this function can be obtained by lengthy, direct
calculations. We will propose a different approach that makes use of
a partition function:
\begin{equation}
\cZ_{AB}=\int d\mu(\rho_A)  \exp\left(-\beta R_{AB}\right),
\end{equation}
where $\beta$ is a fictitious temperature. This approach
is easily generalizable to any other measure of entanglement.
The fictitious
temperature in the partition function (which is the generating
function of the purity) is a ``tool" to fix the value of the purity
(and thus of entanglement). In particular for $\beta =0$ one obtains
typical states, while for larger values of $\beta$ one gets more
entangled states (for $\beta\to\infty$ maximally entangled states).

Henceforth we willassume $N\gg 1$. We will analyze in detail the case $\mu=0$ and then give the results for $M-N=\mu N>0$. Our
problem has been translated into the study of random (reduced) density
matrices $\rho_A$ with
\begin{equation}
\cZ_{AB}=\int_{\lambda_i>0} d^N\lambda\prod_{i<j}(\lambda_i-\lambda_j)^2\delta(1-\sum_{i=1}^N\lambda_i)e^{-\beta N^3 \sum_i\lambda_i^2}.
\end{equation}
As a first step, we introduce a Lagrange multiplier for the delta function
\begin{eqnarray}
\cZ_{AB}&=&N^2 \int\frac{d\xi}{2\pi}\int_{\lambda_i>0} d^N\lambda
\nonumber\\
& & \times e^{i N^2 \xi(1-\sum_i\lambda_i)-\beta N^3 \sum_i\lambda_i^2+2\sum_{i<j}\ln|\lambda_i-\lambda_j|}. \;
\end{eqnarray}
By assuming $N$ large we can look for the stationary point of the
exponent with respect to both the $\lambda_i$'s and $\xi$. The
contour of integration for $\xi$ lies on the real axis but we will
soon see that the saddle point for $\xi$ lies on the imaginary $\xi$
axis. It is then understood that the contour needs to be deformed to
pass by this point parallel to the line of steepest descent.  The
saddle point equations are
\begin{eqnarray}
\label{eq:stat1}
& & -2\beta N^3 \lambda_i+2\sum_{j\neq i}\frac{1}{ \lambda_i-
\lambda_j}-iN^2 \xi=0, \\
& &
\sum_i\lambda_i=1.
\label{eq:normal}
\end{eqnarray}
In the limit of large $N$, by  adopting the natural scaling
\begin{equation}
\lambda_i=\frac{1}{N} \lambda (x_i ), \qquad 0< x_i=\frac{i}{N}\leq 1,
\end{equation}
we can write Eq.\ (\ref{eq:stat1}) as
\begin{equation}
-\beta\lambda + \pvint_0^\infty d\lambda'\frac{\rho(\lambda')}{\lambda-\lambda'}- i\frac{\xi}{2}=0,
\label{eq:sadpt2}
\end{equation}
where
\begin{equation}
\rho(\lambda)=\int_0^1 dx \; \delta(\lambda-\lambda(x))
\label{eq:rhodef}
\end{equation}
is the density of eigenvalues. A similar equation, restricted at $\beta=0$, was studied by Page \cite{page}.

\begin{figure}[htbp]
\centering
\includegraphics[width=\columnwidth]{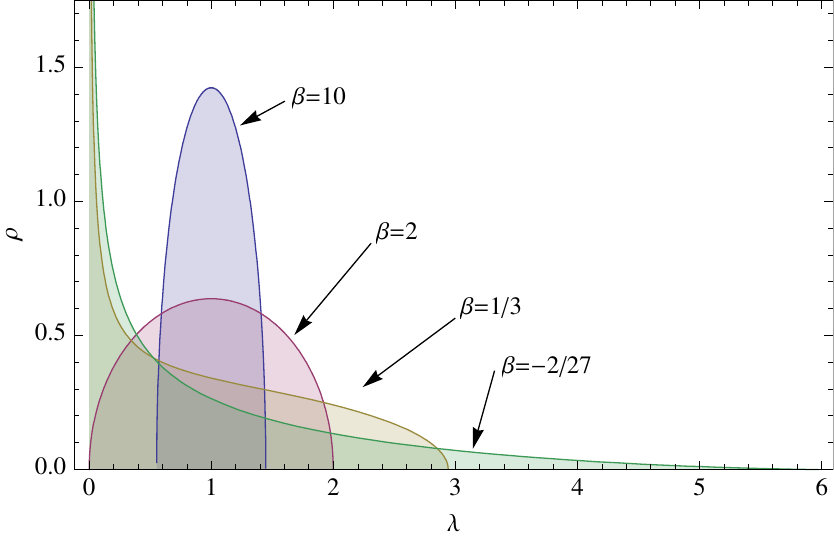}
\caption{Density of eigenvalues at different temperatures.
The phase transitions occur at $\beta_+=2$ and at $\beta_-=-2/27$. }
\label{fig:doev}
\end{figure}

We start at high temperatures $\beta\ll 1$ and assume a solution of
the form \footnote{There exists another solution to Eq.\
(\ref{eq:sadpt2}), which corresponds to reflecting the distribution
around the center of the support $\lambda=a/2$. This however has
higher $F$ than the one studied in the following. We will come back
later to discussing the role of this `parity' symmetry.} (see Fig.\
\ref{fig:doev})
\begin{equation}
\label{eq:ansatz1}
\rho(\lambda)=\frac{\beta}{\pi}\left(\frac{b}{2}+\lambda\right)\sqrt{\frac{a-\lambda}{\lambda}},
\end{equation}
for $0\leq\lambda\leq a$ and $0$ otherwise. This form satisfies the
integral equation as can be promptly verified. The Lagrange
multiplier $\xi$ is related to the parameters $a,b$ by
$\xi=i \beta(a-b)$, and it is purely imaginary, as anticipated.

We can find $a,b$ by imposing normalization and the constraint,
which derive from  (\ref{eq:rhodef})  and (\ref{eq:normal}),
\begin{eqnarray}
\label{eq:normconstr}
\int_0^a d\lambda\rho(\lambda)=1, \qquad
\int_0^a d\lambda\rho(\lambda)\lambda=1.
\end{eqnarray}
By imposing the form (\ref{eq:ansatz1}) we find
\begin{eqnarray}
\frac{\beta}{8} a( a+2 b)=1, \qquad
\frac{\beta}{16} a^2( a+ b)=1 .
\end{eqnarray}

For $\beta_-<\beta<\beta_+$
with
\begin{equation}
\beta_-=-2/27, \quad \beta_+=2 ,
\end{equation}
there is a unique solution of these equations that yields real,
positive $\rho(\lambda):$
\begin{eqnarray}
\label{eq:adibeta}
a(\beta)= \sqrt{\frac{8}{3\beta}}\left(\Delta-\frac{1}{\Delta}\right)
,\quad b(\beta)=\frac{4}{\beta a}-\frac{a}{2},
\end{eqnarray}
where $\Delta=(\sqrt{-\beta/\beta_-}+\sqrt{1-\beta/\beta_-})^{1/3}$.
Notice that
\begin{eqnarray}
& &a(\beta)\sim 4 - 8\beta, \qquad \;\, b(\beta)\sim \beta^{-1} - 4 \beta,
\quad\;\, \mathrm{for}\; \beta\to 0,
\nonumber\\
& &a(\beta)\sim 2 + b(\beta), \quad\;\;\; b(\beta)\sim (\beta_+-\beta)/4,
\;\; \mathrm{for}\; \beta\uparrow \beta_+,
\nonumber\\
& &a(\beta)\sim 18+b(\beta),
\qquad\qquad\qquad\quad\;\;\;  \mathrm{for}\; \beta\downarrow \beta_-,\nonumber\\
& &b(\beta)\sim -12 - \sqrt {12 \left(1-\beta/\beta_-\right)} ,
\qquad \mathrm{for}\; \beta\downarrow \beta_-.
\label{eq:values}
\end{eqnarray}
The average purity is given by
\begin{equation}
\label{eq:puri1}
\langle \pi_{AB} \rangle =\frac{R}{N^3}=\sum_i\lambda_i^2=\frac{1}{N}\frac{\beta}{128} a^3 (5 a+4 b) .
\end{equation}
By using (\ref{eq:values}) one shows that $R(\beta=0)=2 N^2$,
$R(\beta_+)=5 N^2/4$ and $R(\beta_-)=9 N^2/4$ (see later for the
significance of this values).

One can also compute the free energy
\begin{equation}
F=R-\frac{2 N^2}{\beta} \int_0^1 dx \int_0^x dy \log |\lambda(x)-\lambda(y)|
\label{eq:Fdef}
\end{equation}
and using the saddle point equations (\ref{eq:sadpt2}) it is possible to show that
\begin{eqnarray}
& &\int d\lambda \rho(\lambda)\int d\lambda' \rho(\lambda')\log |\lambda'-\lambda|
\nonumber\\
& & \qquad\qquad =\int
d\lambda\rho(\lambda)\left(\log\lambda+\beta\frac{\lambda^2}{2}+i
\frac{\xi}{2}\lambda\right), \;
\label{eq:trick}
\end{eqnarray}
where we also used (\ref{eq:normconstr}), and obtain
\begin{equation}
\frac{F}{N^2}=\frac{1}{8} (6-a) a-\frac{2+a\log (a/4)}{a \beta }+\frac{3 a^4 \beta
}{256},
\end{equation}
in terms of the function $a(\beta)$ introduced above.

Notice that $\beta F$ is the generating function for the connected
correlations of $R$. The radius of convergence in the expansion
around $\beta=0$ defines the behavior of the late terms in the
correlations.

One can find the values of all the cumulants of $R, \pi_{AB}$ (or
connected correlations, the derivatives of $\log \cZ_{AB}$) in the
unbiased distribution at $\beta=0$, when
$\rho(\lambda)=(1/2\pi)\sqrt{(4-\lambda)/\lambda}$. One starts by observing that a series expansion of (\ref{eq:adibeta}) yields
\begin{equation}
a(\beta)=\sum_{l\geq
0}4^{l+1}3^{1-3l}\frac{(3l-1)!}{(2l+1)!(l-1)!} \left(\frac{\beta}{\beta_-}\right)^l.
\end{equation}
By making use of this expression one finds
\begin{equation}
\label{eq:cumuln}
\avg{\avg{\pi_{AB}^{n}}}=
- \frac{(-1)^n}{N^{3n}}\frac{\partial^{n}}{\partial \beta^n}(\beta
F)\big|_{\beta\to 0} = \frac{2^{n+1}}{N^{3n-2}}\frac{(3n-3)!}{(2n)!}.
\end{equation}
The first three cumulants are of course the large-$N$ limits of
known results \cite{aaa} (for small $N$ exact expressions for the first 5 cumulants can be also found in \cite{Giraud}).

\begin{figure}[htbp]
\centering
\includegraphics[width=\columnwidth]{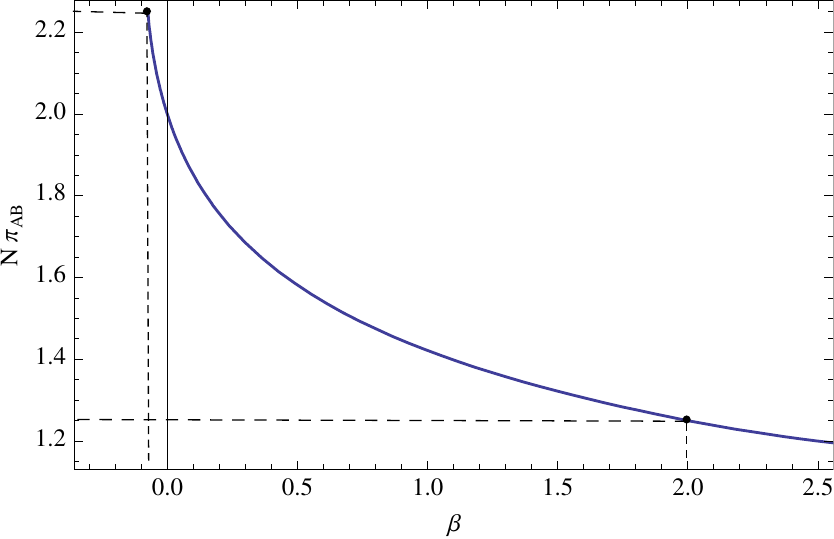}
\caption{$\langle\pi_{AB}\rangle$ as a function
of the inverse temperature. Notice the value
$\langle\pi_{AB}\rangle=2/N$ at $\beta=0$ (typical states). In the
$\beta\to\infty$ limit we find the minimum
$\langle\pi_{AB}\rangle=1/N$. The phase transitions described in the
text are at $\beta_-=-2/27,\langle\pi_{AB}\rangle=9/4N$ (left point)
and $\beta_+=2,\langle\pi_{AB}\rangle=5/4N$ (right point).}
\label{pi}
\end{figure}

We are now ready to unveil the presence of two phase transitions.
The most evident one is at the end of the radius of convergence of
the small $\beta$ expansion, which occurrs at $\beta_-$. We
can extend our equations smoothly down to $\beta_-$ but
not below. At $\beta_-$ we have
$\rho(\lambda)=2/(27\pi) (6-\lambda)^{3/2}/\sqrt{\lambda}$ and
 $\pi_{AB}=9/4N$ (see Figures \ref{fig:doev} and
\ref{pi}). The derivative at the right edge of eigenvalue
density vanishes and some eigenvalues can evaporate to
$+\infty$.\footnote{It is likely that for arbitrarily small
and negative $\beta$ this phase is unstable for
non-perturbative effects to an almost separable phase where, say, $\lambda_1=1-\Ord{1/N}$
and $\lambda_{n>1}=\Ord{1/N^2}$. The radius of convergence of the
series expansion of $F(\beta)$ for $\beta\to 0$ is however blind to such
non-perturbative effects.} The limits $\beta\to\beta_-$ and
$N\to\infty$ can be combined (double-scaling limit) to interpret the
free energy as the partition function of random 2-D surfaces (a
theory of pure gravity). Using (\ref{eq:cumuln}) we see that around
$\beta_-$ the free energy $F\propto (\beta-\beta_-)^{5/2}+\rm{less\
singular}$ \cite{Di Francesco:1993nw}. In fact,
if one relaxes the unit trace condition, our partition function
$\cZ$ has been studied in the context of random matrix theories
\cite{Morris91}. The objects generated in this way
correspond to chequered polygonations of surfaces. Our calculations
show that the constraint $\Tr\ \rho_A=1$ is irrelevant for the
critical exponents.

The other phase transition occurs as $\beta$ is increased (the
temperature decreased). The value of $b$ decreases continuously and
eventually vanishes at $\beta_+$ (where $\pi_{AB}=5/4N$), becoming
$b<0$ for $\beta>\beta_+$. The solution (\ref{eq:ansatz1}) is not valid
anymore, since $\rho(\lambda)$ becomes negative for $\lambda<-b/2$.
We have to look for another solution, and, by noting that at
$\beta_+$, $\rho(\lambda)=(\beta_+/\pi)\sqrt{\lambda(2-\lambda)}$ (see
Fig.\ \ref{fig:doev}), we do so in the usual semicircle form
\begin{equation}
\label{eq:semicircle}
\rho(\lambda)=\frac{\beta}{\pi}\sqrt{\lambda-b}\sqrt{a-\lambda}.
\end{equation}

The normalization and the constraint yield
\begin{eqnarray}
\frac{\beta}{8} ( a-b)^2=1, \qquad
\frac{\beta}{16} ( a-b)^2 (a+b)=1 .
\end{eqnarray}
This can be easily solved to find
\begin{equation}
a=1 +\sqrt{\frac{\beta_+}{\beta}}, \qquad b=1-\sqrt{\frac{\beta_+}{\beta}}
\end{equation}
and hence
\begin{equation}
R=N^2\left(1+\frac{1}{2\beta}\right).
\end{equation}
Moreover, from (\ref{eq:Fdef})-(\ref{eq:trick}), one gets
\begin{equation}
\frac{F}{N^2}=1+\frac{3}{4\beta} +\frac{1}{2\beta}\log(2\beta).
\end{equation}
We can now notice how the phase transition at $\beta_+$ is due to
the restoration of a ${\mathbb Z}_2$ symmetry $P$ (`parity') present
in Eq.\ (\ref{eq:sadpt2}), namely the reflection of the distribution
$\rho(\lambda)$ around the center of its support ($\lambda=a/2$ for
$\beta\leq\beta_+$ and $1$ for $\beta>\beta_+$). For $\beta\leq \beta_+$ there are two
solutions linked by this symmetry, and we picked the one with the
lowest $F$; at $\beta_+$ this two solutions coincide with the
semicircle (\ref{eq:semicircle}), which is invariant under $P$ and
becomes the valid and stable solution for higher $\beta$.

One can also determine the expression for the entropy $S=\beta(R-F)$,
which counts the number of states with a given value of the purity.
The expression for $\beta<\beta_+$ is quite involved and we will not write
it here, while for $\beta\geq \beta_+$ it is easy to see that:
\begin{equation}
\frac{S}{N^2}=-\frac{1}{4}-\frac{1}{2}\log(2\beta), \quad \beta\geq\beta_+.
\end{equation}
In the critical region, $\beta \to \beta_+$, we find
\begin{eqnarray}
\frac{S}{N^2} \sim -\frac{1}{4}-\log 2 -\frac{\beta-\beta_+}{4} + \theta(\beta-\beta_+) \frac{(\beta-\beta_+)^2}{16},\;
\end{eqnarray}
where $\theta$ is the step function. We see that $S$ is continuous
at the phase transitions, together with its first derivative
although the second derivative is discontinuous. So this is a second
order phase transition.

Notice that the entropy is unbounded from below when $\beta \to
+\infty$. The interpretation of this result is quite
straightforward: the minimum value of $\pi_{AB}$ is reached on a
sub-manifold (isomorphic to $SU(N)/Z_{N}$
\cite{Kus01}) of dimension $N^2-1$, as opposed to the typical case
vectors which form a manifold of dimension $2N^2-N-1$ in the
Hilbert space $\mathcal{H}$. Since this manifold has zero
volume in the original Hilbert space, the entropy, being the
logarithm of this volume, diverges.

With the same techniques, starting from (\ref{eq:measure}) we can
find the cumulants of the purity for unbalanced bipartitions.
Leaving the details for a forthcoming publication we report the
results for the first five cumulants only:
\begin{eqnarray}
\avg{\pi_{AB}}&=&\frac{1}{N}\frac{2+\mu}{1+\mu},\quad
\avg{\avg{\pi_{AB}^2}}=\frac{1}{N^4}\frac{2}{(1+\mu)^2},\nonumber\\
\avg{\avg{\pi_{AB}^3}}&=&\frac{8}{N^7}\frac{2+\mu}{(1+\mu)^4},\quad
\avg{\avg{\pi_{AB}^4}}=\frac{48}{N^{10}}\frac{6+6\mu+\mu^2}{(1+\mu)^6},\nonumber\\
\avg{\avg{\pi_{AB}^5}}&=&\frac{384}{N^{13}}\frac{22+33\mu+13\mu^2+\mu^3}{(1+\mu)^8}.
\label{eq:cumul2}
\end{eqnarray}
where $\mu=(M-N)/N$. For $\mu=0$ these reduce to
the results of the previous section.

{\bf Conclusions.}
We have calculated the generating function of a typical entanglement
measure, averaged over the Hilbert space. We have shown that, when
interpreted as a partition function, it possesses multiple phase
transitions. In the different phases the distribution of Schmidt coefficients have different profiles. Sudden changes of these profiles occur at the phase transitions.

We have studied these phase transition(s) as a function of a fictitious
temperature $\beta$, introduced to define the generating function of
the purity. This fictitious temperature can also be thought of as
localizing the measure on set of states with entanglement larger or
smaller than the typical one \cite{Kus01} (in the same way
temperature is used in classical statistical mechanics to fix the
energy to a given value in the thermodynamic limit).

Notice that the phase transitions investigated here, that appear in
the study of the generating functions of any entanglement measure,
are not quantum phase transitions (QPT). Since entanglement is known
to be a good indicator of QPTs \cite{qpt}, it would be interesting
to investigate the link, if any, between these different
transitions. 

In conclusion, by using techniques borrowed from the study of random
matrix theory, we gave a complete characterization of the
statistics of one entanglement measure.  We also proposed one direction in which random matrix theory is likely to play a
significant role in the study of entanglement, namely the role of the phase transitions found in random matrix theory as describing the change in the profile of typical, less or more entangled states.

{\bf Acknowledgements} We thank G.~Marmo for discussions. A.S. would
also like to thank him for his hospitality in Napoli, where part of
this work has been completed. This work is partly supported by the
European Community through the Integrated Project EuroSQIP.


\begin{thebibliography}{99}

\bibitem{woot}
W. K. Wootters, Quantum Inf. and Comp., \textbf{1}, 27 (2001); L.
Amico, R. Fazio, A. Osterloh and V. Vedral ``Entanglement in
Many-Body Systems," arXiv:quant-ph/0703044 (Rev. Mod. Phys., in
print).

\bibitem{multipart}
V. Coffman, J. Kundu and W. K. Wootters, Phys. Rev. A \textbf{61},
052306 (2000); A. Wong and N. Christensen, Phys. Rev. A \textbf{63},
044301 (2001); D. Bruss, J. Math. Phys. \textbf{43}, 4237 (2002);
D.A. Meyer and N.R. Wallach, J. Math. Phys.
\textbf{43}, 4273 (2002).

\bibitem{parisi}
M. Mezard, G. Parisi and M. A. Virasoro, {\it Spin Glass Theory and
Beyond} (World Scientific, Singapore, 1987).


\bibitem{FFPP}
P.~Facchi, G.~Florio, G.~Parisi, S.~Pascazio, arXiv:0710.2868v1
[quant-ph].

\bibitem{nielsen}
 M.A. Nielsen and I.L. Chuang,
    {\it Quantum Computation and Quantum Information}
    (Cambridge University Press, Cambridge, 2000).

\bibitem{aaa}
E.\ Lubkin, J.\ Math.\ Phys. \textbf{19}, 1028 (1978); S. Lloyd and
H. Pagels, Ann. Phys., NY, \textbf{188}, 186 (1988); K
\.{Z}yczkowski and H.-J. Sommers, J. Phys. A \textbf{34}, 7111
(2001); A. J. Scott and C. M. Caves, J. Phys. A: Math. Gen.
\textbf{36}, 9553 (2003).

\bibitem{Giraud}
O. Giraud, J.\ Phys.\ A: Math.\ Theor.\ {\bf 40} (2007) 2793.

\bibitem{page}
D.N. Page, Phys. Rev. Lett. \textbf{71},
1291 (1993).

\bibitem{Di Francesco:1993nw}
  P.~Di Francesco, P.~H.~Ginsparg and J.~Zinn-Justin,
  Phys.\ Rept.\  {\bf 254}, 1 (1995)
  [arXiv:hep-th/9306153].

\bibitem{Majumdar07}
S. N.\ Majumdar, O.~Bohigas, A.~Lakshminarayan, preprint:
arXiv:0711.0677v1 [cond-mat.stat-mech].

\bibitem{Morris91} T.~R.~Morris, Nuclear
Physics {\bf B 356}, 703 (1991).

\bibitem{Kus01}
    M.~M.\ Sinolecka, K.\ Zyczkowski, M.\ Kus,
    Acta Physica Polonica B: {\bf 33},  p.~2081 (2001).

\bibitem{qpt}
A. Osterloh, L. Amico, G. Falci, and R. Fazio, Nature (London)
\textbf{416}, 608 (2002); T. J. Osborne and M. A. Nielsen, Phys. Rev. A
\textbf{66}, 032110 (2002); L.-A.\ Wu, M.~S.\ Sarandy,
D.~A.\ Lidar, \prl {\bf 93}, 250404 (2004); D.\ Larsson, H.\
Johannesson, \pra, {\bf 73}, 042320 (2006). 

\end{thebibliography}
\end{document}